\documentclass[review]{elsarticle}

\usepackage{lineno ,hyperref}
\modulolinenumbers[5]

\journal{Journal of \LaTeX\ Templates}

\begin{document}

\begin{frontmatter}

\title{Generation of Synthetic Rat Brain MRI scans with a 3D Enhanced Alpha-GAN}

\author[mymainaddress]{André Ferreira \corref{mycorrespondingauthor}}
\ead{a81350@alunos.uminho.pt}
\author[mysecondaryaddress]{Ricardo Magalhães}
\author[mysecondaryaddress]{Sébastien Mériaux }

\author[mymainaddress]{Victor Alves}

\cortext[mycorrespondingauthor]{Corresponding author}

\address[mymainaddress]{Centro Algoritmi, University of Minho, Braga, Portugal}
\address[mysecondaryaddress]{Université Paris-Saclay, CEA, CNRS, BAOBAB, NeuroSpin, 91191 Gif-sur-Yvette, France }

\fntext[]{MRI (Magnetic Resonance Imaging)}
\fntext[]{GAN (Generative Adversarial Networks)}
\fntext[]{DL (Deep Learning)}
\fntext[]{VAE (Variational AutoEncoder)}
\fntext[]{SN (Spectral Normalisation)}
\fntext[]{GDL (Gradient Difference Loss )}
\fntext[]{GM (Grey Mater), WM (White Mater), CSF (Cerebrospinal Fluid)}

\begin{abstract}
Translational brain research using Magnetic Resonance Imaging (MRI) is becoming increasingly popular as animal models are an essential part of scientific studies and more ultra-high-field scanners are becoming available. Some disadvantages of MRI are the availability of MRI scanners and the time required for a full scanning session (it usually takes over 30 minutes). Privacy laws and the 3Rs ethics rule also make it difficult to create large datasets for training deep learning models. Generative Adversarial Networks (GANs) can perform data augmentation with higher quality than other techniques. In this work, the alpha-GAN architecture is used to test its ability to produce realistic 3D MRI scans of the rat brain. As far as the authors are aware, this is the first time that a GAN-based approach has been used for data augmentation in preclinical data. The generated scans are evaluated using various qualitative and quantitative metrics. A Turing test conducted by 4 experts has shown that the generated scans can trick almost any expert. The generated scans were also used to evaluate their impact on the performance of an existing deep learning model developed for segmenting the rat brain into white matter, grey matter and cerebrospinal fluid. The models were compared using the Dice score. The best results for whole brain and white matter segmentation were obtained when 174 real scans and 348 synthetic scans were used, with improvements of 0.0172 and 0.0129, respectively. Using 174 real scans and 87 synthetic scans resulted in improvements of 0.0038 and 0.0764 for grey matter and CSF segmentation, respectively. Thus, by using the proposed new normalisation layer and loss functions, it was possible to improve the realism of the generated rat MRI scans and it was shown that using the generated data improved the segmentation model more than using the conventional data augmentation.
\end{abstract}

\begin{keyword}
\texttt{Alpha Generative Adversarial Network \sep Data Augmentation \sep Synthetic Data \sep MRI Rat brain}

\end{keyword}

\end{frontmatter}


\section{Introduction}

Translational MRI research of the brain is becoming very popular and rodents are often used as starting points for various complex studies, e.g. in the field of neuroscience. The ability to obtain multimodal information in the same animal (e.g. anatomy, function, metabolism) makes MRI a powerful imaging modality for neuroscience research in humans, as it allows non-invasive in vivo studies from disease diagnosis to treatment monitoring, as it allows the study of anatomy as well as functions and metabolism, and has no adverse effects \cite{denic2011mri,Brockmann2007}.

The use of MRI allows a simplified transfer of results from animal models to humans, as the same acquisition scheme can be used to generate datasets in both humans and animals. Although it is known that there are differences between species and the parameters measured are not identical due to structural and functional differences, some homologies may be significant to allow translation. The use of rodents remains essential to study the brain in physiological or pathological states to develop new brain models or new therapeutic strategies. The development of specific rodent models makes it possible to work under more controlled conditions and to go beyond what is possible in human studies \cite{denic2011mri, Brockmann2007, Barriere2019}.

A variety of preclinical studies have already shown the importance of using rodents, e.g. Ricardo Magalhães et al. (2019) \cite{Magalhaes2018, Magalhaes2018a, Magalhaes2019} studied the structural and functional modifications of rat brain induced by the exposure to stress, Boucher et al. (2017) \cite{Boucher2017} investigated in glioblastoma-bearing mice the possibility of using genetically tailored magnetosomes as MRI probes for molecular imaging of brain tumours, Vanhoutte et al. (2005) \cite{Vanhoutte2005} used MRI and mice to detect amyloid plaques for the detection of Alzheimer's disease in vivo.

Although the use of rodents has many advantages, MRI acquisitions on small animals require special scanners with a high magnetic field and special high-frequency coils to ensure an optimal signal-to-noise ratio and thus high spatial and temporal resolution. However, this is becoming more and more of a reality \cite{denic2011mri,Brockmann2007}. The relative lack of suitable methods for data processing is another limitation that has resulted in the slow growth of the use of MRI to study the rodent brain \cite{Magalhaes2018}. 

One of the disadvantages of MRI is the time needed to perform a scan with good resolution and satisfactory data quality. A typical MRI session usually takes more than 30 minutes and up to an hour. Availability of scanners is also a problem, as they are not always operational for use or their use is restricted by government regulations. Data protection laws can also lead to a lack of data \cite{Foroozandeh2020}. For preclinical experiments, the 3Rs ethics rule \cite{russell1959principles} limits the number of animals scanned, which reduces the number of scans available. In addition, the MRI scanner is very expensive and so is each scan. Some efforts have been made to overcome these problems, but the cost per available data ratio remains too high.

Machine learning is on the rise in the field of medical imaging. These techniques can be used to train various machine learning models such as Deep Learning (DL) models to assist specialists in image processing (e.g. segmentation), disease detection, decision support and other tasks. The main limitation is that they usually require a huge amount of data to train properly and perform well.

To mitigate this problem, conventional data augmentation \cite{Nalepa2019} and generative models have been used. These techniques have improved the ability of DL models to achieve the goals for which they were developed, but only slightly \cite{Sandfort2019, Motamed2020}. Conventional data augmentation does not fill all existing gaps in the dataset, and some generative models, e.g. AutoEncoder and Variational AutoEncoder (VAE), cannot produce sufficiently realistic scans, and the scans are characterised by blurriness and low detail \cite{el2019deep}. To deal with this situation, Goodfellow et al. (2014) introduced a new approach to improve generative models, called Generative Adversarial Networks \cite{Goodfellow2014}. 

\begin{figure}[h]
    \centering
    \includegraphics[width=1\textwidth]{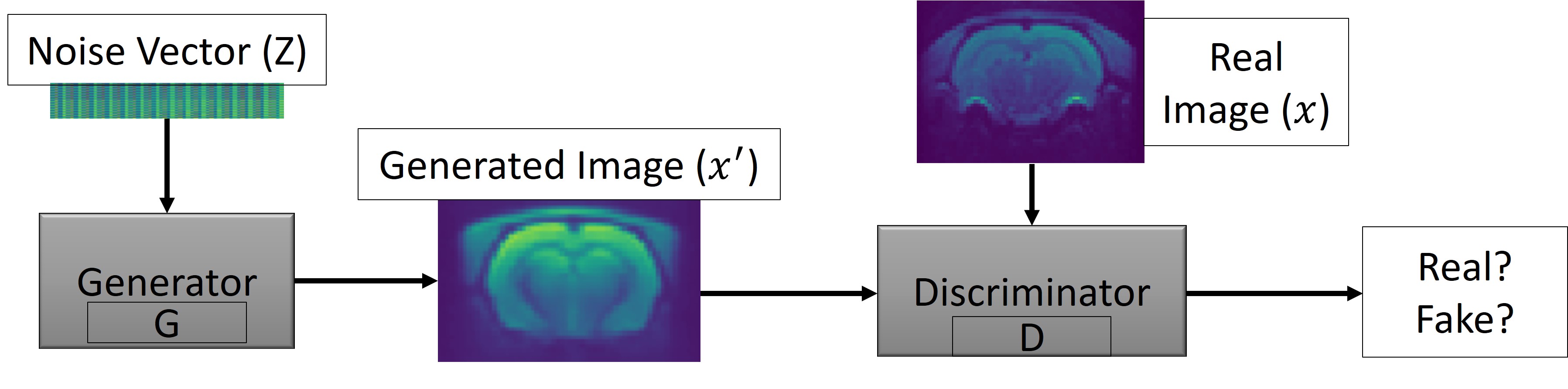}
    \caption{Representation of the basic GANs architecture with noise vector (z), generator (G), synthetic data (x’), real data (x) and discriminator (D).}
    \label{fig1}
\end{figure}

GANs are a type of deep neural architecture with two simultaneously trained networks, one generating the image and the other specialised in discrimination. The first aims to generate fake images from an input vector (usually a Gaussian distribution) and to fool the discriminator, which is a classifier that evaluates whether the images generated by the generator are realistic. The discriminator gives a probability of truth each time it receives an image, with higher values corresponding to images that are closer to reality. A probability close to 0.5 is the optimal solution, as this means that the discriminator is not able to distinguish the real images from the fake ones \cite{Alqahtani2019, Gui2020}. The whole process is illustrated in Figure \ref{fig1}.

Given the results that GANs have achieved in various tasks \cite{Brock2019, Karras2020}, it will be possible to disseminate a larger dataset than the original one without protection restrictions once GANs have the ability to anonymise \cite{shin2018medical}. In addition, data augmentation with GANs is more robust than traditional data augmentation alone. This reduces the amount of real MRI scans needed to train good medical DL models or even improve some existing models. The fact that the entire 3D MRI volume is used increases the training difficulty exponentially, but by using the entire scans it is possible to save time and resources by reducing the number of samples required.

The use of MRI for this study is advantageous from a practical and ethical point of view, as high-resolution images can be obtained in a non-invasive way, reducing animal suffering and allowing them to be observed over a longer period. As far as the authors are aware, the use of GANs in preclinical research is a little addressed topic, but necessary as it is fully in line with the ethical 3R rule by reducing the number of scanning sessions.

\section{Material and methods}
\subsection{Sigma dataset of rat brains}
\begin{figure}[h]
    \centering
    \includegraphics[width=1\textwidth]{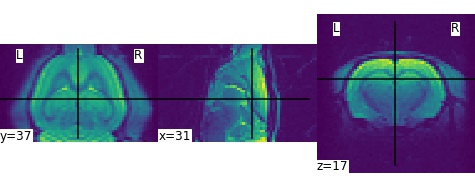}
    \caption{Example of one functional MR image from the Sigma rat brain dataset displayed in all three planes (coronal, sagittal and axial planes, respectively).}
    \label{fig2}
\end{figure}

An MRI dataset of the Wistar rat brain from the Sigma project was used to test the ability of GANs to produce synthetic scans. A total of 210 scanning sessions were performed using a Bruker preclinical ultra-high field scanner (11.7 Tesla) and a 4x4 surface coil dedicated to the rat head. A T2-weighted Echo Planar Imaging sequence was implemented to acquire resting-state functional data, with a spatial resolution of 0.375mm x 0.375mm x 0.5mm over a matrix of 64x64x40, a TR of 2000ms, a TE of 17.5ms and 9 averages.  Figure \ref{fig2} shows slices from three different planes (i.e. coronal, sagittal and axial) of a functional MR image of the rat brain. The dataset consists of 210 scans with a resolution of 64x64x40, which have been pre-processed to avoid complex values. For more information on the rat brain sigma dataset, see Barrière et al. (2019) or Magalhães et al. (2018) \cite{Barriere2019, Magalhaes2018a}. 

In this work, a different number of scans and sources (real or synthetic) were used in some steps of the experiment. Therefore, each dataset is formally defined as  D\textsubscript{s}\textsuperscript{N} =  {(x\textsubscript{i}, y\textsubscript{i})}\textsubscript{i=1}\textsuperscript{N}, where s indicates whether the scans are synthetic (s) or real (r), x the scans, y the respective labels and N the number of scans. The original sigma dataset of rat brains is formally defined as  D\textsubscript{r}\textsuperscript{210} =  {(x\textsubscript{i}, y\textsubscript{i})}\textsubscript{i=1}\textsuperscript{210}.

\subsection{Overall process workflow}

This work was developed in Python, drawing on two important DL frameworks, PyTorch \cite{NEURIPS2019_9015} and MONAI \cite{Monai}. The specifications of the workstations are given in Table \ref{tab1}.

\begin{table}[h]
\centering
\small
\caption{Workstation Specifications}
\begin{tabular}{|l|l|}
\hline
Operative   System & Ubuntu 18.04.3 LTS (64   bits)                                                                                                                      \\ \hline
CPU                & Intel Xeon E5-1650                                                                                                                                  \\ \hline
GPU                & \begin{tabular}[c]{@{}l@{}}GPU – NVIDIA P6000\\ Cuda Parallel-Processing   Colors 3840\\ 24 GB GDDR5X\\ FP32 Performance 12 TFLOPS\end{tabular} \\ \hline
Primary   Memory   & 64Gb                                                                                                                                                \\ \hline
Secondary   Memory & \begin{tabular}[c]{@{}l@{}}2 Disks of 2TB\\ 1 Disk of 512Gb\end{tabular}                                                                            \\ \hline
\end{tabular}
\label{tab1}
\end{table}

\begin{figure}[h!]
    \centering
    \includegraphics[width=0.91\textwidth]{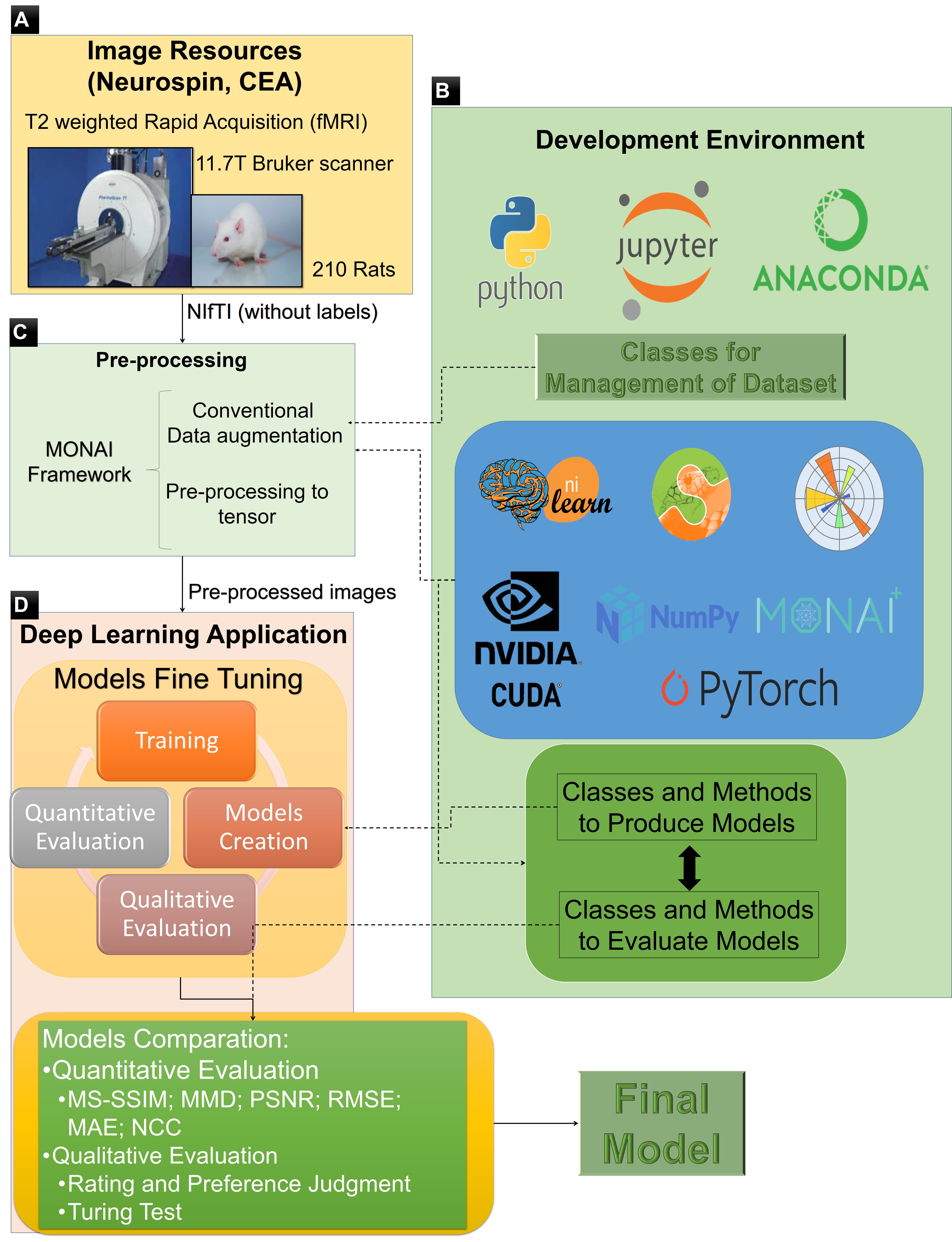}
    \caption{Overall training and evaluation process workflow.}
    \label{fig3}
\end{figure}

The entire process flow of training and assessment is described in Figure \ref{fig3}. Figure \ref{fig3}-A, the Image Resources block represents the acquisition of the MRI scans and the creation of the dataset. Figure \ref{fig3}-B, the Development Environment block describes the development environment with all frameworks, libraries and other dependencies for training the models. Figure \ref{fig3}-C, the Preprocessing block corresponds to the preprocessing of the dataset (always working with the whole 3D scan), such as resizing to 64x64x64 with constant padding of zero values, intensity normalisation between -1 and 1, and some conventional data augmentation. Figure \ref{fig3}-D, the Deep Learning application block, is the final step where the training and evaluation of the models are done and where the best model is selected.

Some conventional data augmentations such as zooming, rotating, adding Gaussian noise, flipping, shifting and scaling the intensity were done as this can fill some gaps in the dataset. The size of the input random vector is an important consideration as it should be large enough to represent the dataset but not too large to avoid overfitting. Therefore, the sizes 500/1000 were tested.

\begin{figure}[h]
    \centering
    \includegraphics[width=1\textwidth]{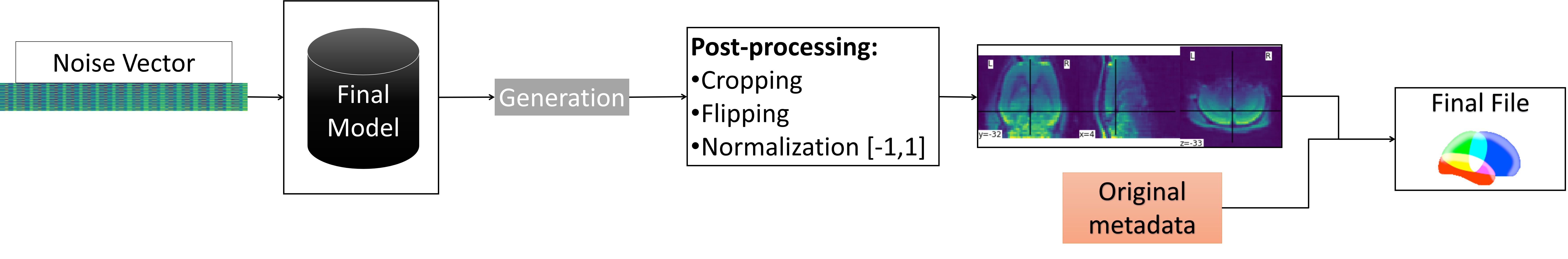}
    \caption{Overall generation workflow.}
    \label{fig4}
\end{figure}

The creation of the final files is the last step, shown in Figure \ref{fig4}. To generate a new scan, a random vector (e.g. a noise vector from a Gaussian distribution) and the final model from the training process are needed. Then the generated scan is subjected to post-processing, e.g. flipping, cropping and normalising, resulting in an image with the same properties as the original file. Finally, it is possible to generate a 3D NIfTI file from the dataset using the generated scan and metadata (headers and affines) from an original file (from the D\textsubscript{r}\textsuperscript{210}).

Eleven models were created, each trained for 200000 iterations with a stack size of 4, but only the two best and the base model for comparison are presented in this article. The first model was the baseline based on the Kwon et al. (2019) model, where generator, discriminator, encoder and code discriminator networks are identical and the latent space is of size 1000 \cite{Kwon2019}. The use of this architecture is more computationally intensive than some traditional GANs architectures because it involves four networks instead of just two. However, this choice was justified because the $\alpha$-GAN architecture avoids mode collapse and fuzziness through the introduction of a VAE and a code discriminator into the GAN network \cite{Kwon2019}. The loss functions used in this training process were (1) and (2) for generator/encoder, (3) for code discriminator and (4) for discriminator with  $\lambda_1=\lambda_2=10$ (based on the experiments of Kwon et al. (2019)):

$L_{GD}=-E{_{z{_{e}}}}[D(G(z_{e}))]-E{_{z{_{r}}}}[D(G(z_{r}))] (1)$

$L_{G1}=L_{GD}-E{_{z{_{e}}}}[C(z_{e})]+ \lambda_{2}||x_{real}-G(z_{e})||_{L1} (2)$

$L_{C}=E{_{z{_{e}}}}[C(z_{e})]-E_{z_{r}}[C(z_{r})]+\lambda_{1}L_{GP-C} (3)$

$L_{D}=-L_{GD}-2E{_{x{_{real}}}}[D(x_{real})]+\lambda_{1}L_{GP-D} (4)$

where L\textsubscript{GD} denotes the feedback from the discriminator, D denotes the Discriminator, G denotes the Generator, C denotes the Code Discriminator, z\textsubscript{e} denotes the latent vector of the encoder, z\textsubscript{r} denotes the input random vector, x\textsubscript{real} denotes a real scan, L\textsubscript{GP-D} denotes the gradient penalty of Discriminator, L\textsubscript{GP-C} denotes the gradient penalty of Code Discriminator, L1 denotes the L1 loss function and $E$ the total distribution. Vertical flipping was the only conventional data augmentation used to train this first model. The Adam optimizer \cite{Kingma2015} was used with a learning rate of 0.0002, betas of 0.9, 0.999 and eps of 10\textsuperscript{-8}. This architecture was called $\alpha$-WGAN\_ADNI. 

\begin{figure}[h]
    \centering
    \includegraphics[width=1\textwidth]{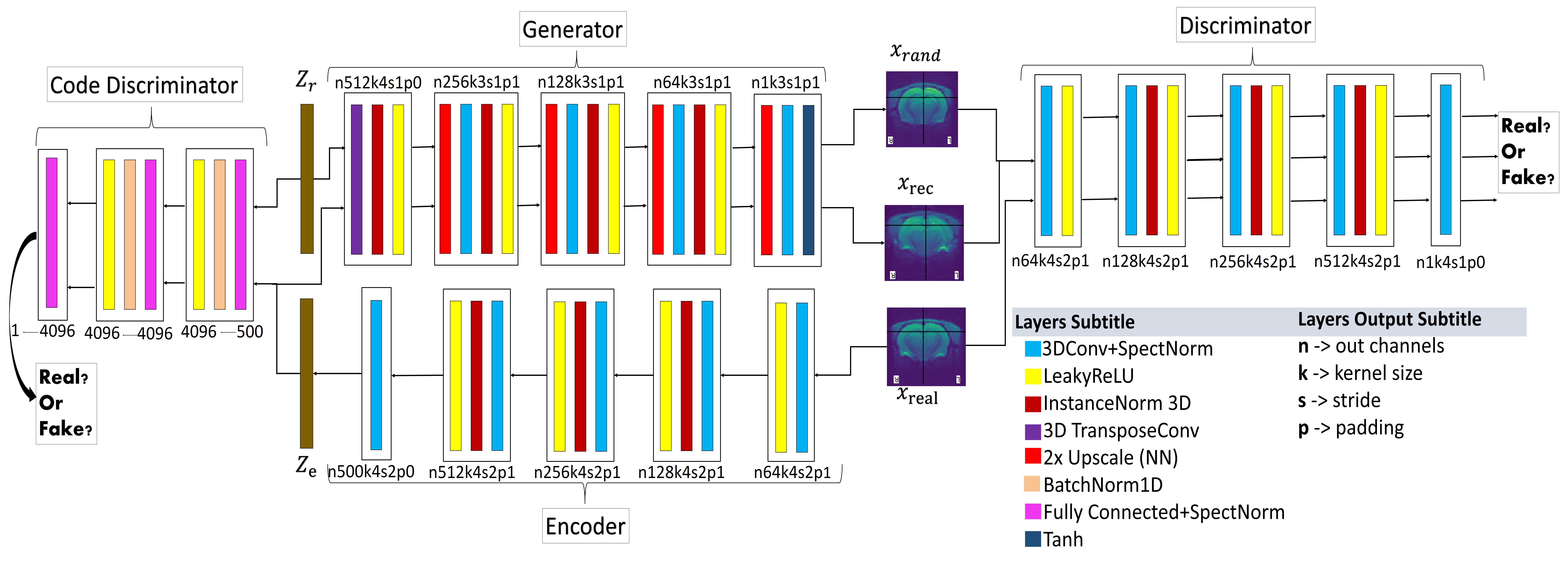}
    \caption{$\alpha$-WGANSigmaRat1 architecture. Near each layer there is a code that explains the result of each convolution, e.g., n256k3s1p1: n256, size of the output channels (in this case 256); k3, kernel size (cubic dimension 3x3x3); s1, stride (1x1x1); p1, padding (1x1x1).}
    \label{fig5}
\end{figure}

The new loss functions and normalisation layer used in this work are described below. The first proposed architecture, based on Sun et al. (2020) \cite{Sun2020} is shown in Figure \ref{fig5} in which Spectral Normalisation (SN) \cite{Miyato2018} was added after each convolution to stabilise the training, especially the training of the discriminator, the batch normalisation layers were replaced by instance normalisation layers to avoid some artefacts, and the activation function LeakyReLU was used instead of ReLU to speed up the training and improve the results \cite{Agrawal2019}. 

This architecture was called $\alpha$-WGANSigmaRat1. The loss functions used to train this model were the same as those used for the $\alpha$-WGAN\_ADNI model, except for the generator, for which a new loss function was introduced (5): 

$L_{G2}=L_{GD}-E{_{z{_{e}}}}[C(z_{e})]+ \lambda_{1}||x_{real}-G(z_{e})||_{L1}+ \lambda_{1}||x_{real}-G(z_{e})||_{MSE} (5)$

For this training process, all previously mentioned conventional data augmentations (zoom, rotation, Gaussian noise, flip, translation and scaling intensity) as well as the Adam optimiser and a random vector size of 500 were used. 

\begin{figure}[h]
    \centering
    \includegraphics[width=1\textwidth]{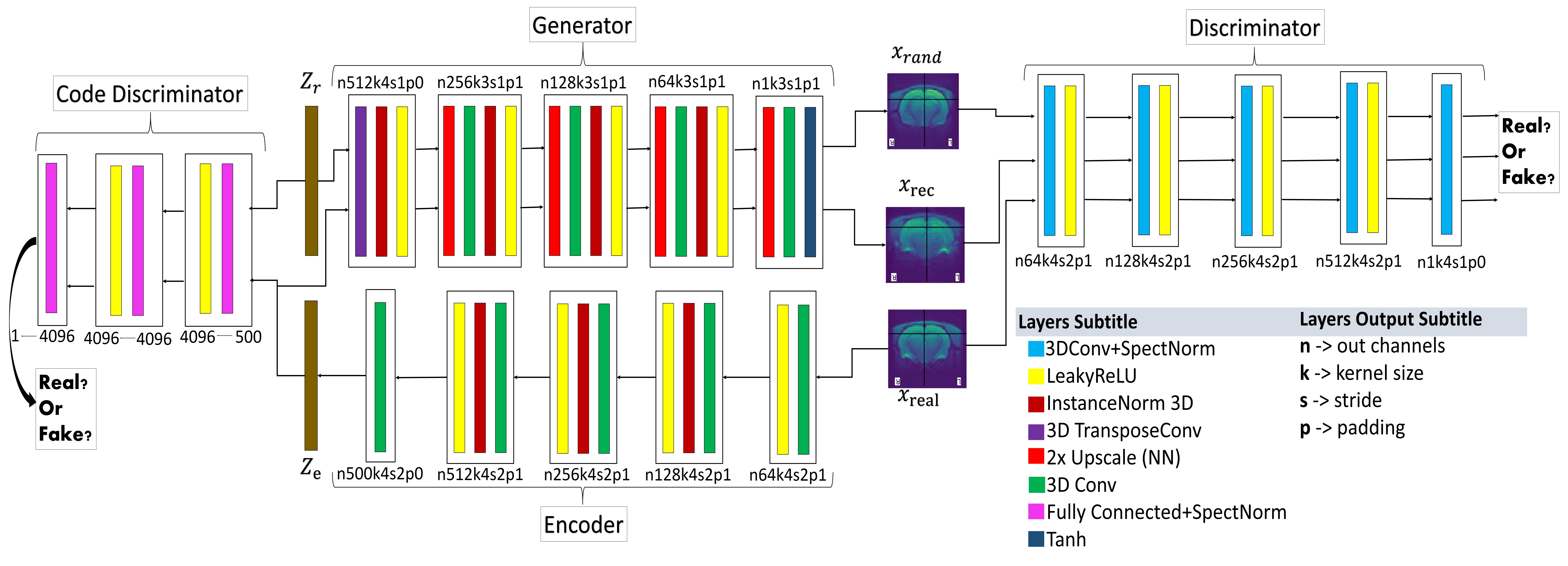}
    \caption{$\alpha$-WGANSigmaRat2 architecture. Near each layer is displayed a code that explains the result of each convolution, e.g., n256k3s1p1: n256, size of the output channels (in this case 256); k3, kernel size (cubic dimension 3x3x3); s1, stride (1x1x1); p1, padding (1x1x1).}
    \label{fig6}
\end{figure}

The last proposed model architecture is shown in Figure \ref{fig6}. The main changes compared to the $\alpha$-WGANSigmaRat1 architecture were the removal of SN after each convolution only in the generator and encoder, since in the original work by Miyato et al. (2018) \cite{Miyato2018} the SN was created to stabilise the training of the discriminator. The instance normalisation layers were also removed in the discriminator and in the code discriminator to avoid artefacts and reduce computational costs. In the new loss function (6), the L1 loss function was replaced by the Gradient Difference Loss (GDL) \cite{Mathieu2016}, which is described in the latest works on super-resolution \cite{Mathieu2016, Chen2018, Sanchez2018}:

$L_{G3}=L_{GD}-E{_{z{_{e}}}}[C(z_{e})]+ \lambda_{3}||x_{real}-G(z_{e})||_{GDL}+ \lambda_{2}||x_{real}-G(z_{e})||_{MSE} (6)$

The loss functions used to train this model were (3) for the code discriminator, (4) for the discriminator and (6) for the generator/encoder with $\lambda_1=\lambda_2=100$ and $\lambda_3=0.01$. The chosen values for  $\lambda_1$, $\lambda_2$ and $\lambda_3$ proved to be more stable than other values after some experimental tests.

All conventional data augmentations were performed, the size of the input random vector was 500, and a new optimiser was used, AdamW \cite{Loshchilov2019} with a learning rate of 0.0002, betas of 0.9, 0.999, eps of 10\textsuperscript{-8} and a weight decay of 0.01. The AdamW is known to have a more stable weight decay than the Adam. This architecture was named $\alpha$-WGANSigmaRat2.

The difference between the synthesis of rat and human MRI brains is the resolution of the input scans. Here, scans with a resolution of 64x64x40 were used, but nowadays (with the arrival of better GPUs) it is possible to process scans with higher resolutions (e.g. 256x256x256). Since an alpha-GAN architecture with new loss functions and a special normalisation (SN) is used, the training can be performed with human MRI scans without much difference, except for the longer runtime.

\section{Results}

As explained in the section Material and methods, eleven different models were trained, but only the baseline model (for comparison) and the two best models are examined and evaluated here.

To select the best metrics for evaluating the different models, various MRI-related articles dealing with 3D scans and GANs were reviewed. The evaluation metrics were then divided into two categories: quantitative (MS-SSIM, MAE, NCC, MMD and Dice score) and qualitative (Visual Turing Test performed by experts). The most commonly used metrics for evaluating GANs, e.g. Frechet Inception Distance, Inception Score and Kernel Inception Distance, were not used due to the complexity of their adaptation to 3D volumes. The mean and standard deviation were calculated for each metric. For the MMD, MAE and NCC metrics, 21000 (number of scans in dataset *100 epochs) comparisons were made; for the MS-SSIM, only 2100 comparisons were made for both real and generated data, as the computational power required was much higher than for the other metrics. The MS-SSIM value was calculated under the same conditions as in Kwon et al. (2019) \cite{Kwon2019}, i.e. with a batch size of 8. The MS-SSIM value should be similar to the MS-SSIM value of the real dataset to prove that the distribution is similar.

\begin{figure}[]
    \centering
    \includegraphics[width=1\textwidth]{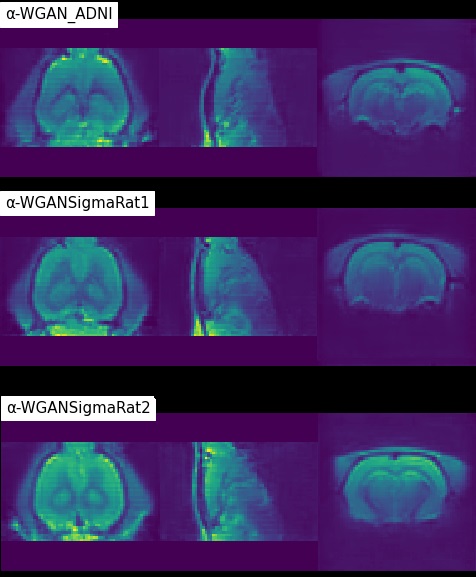}
    \caption{Coloured visualization of scans generated by different models in the coronal, sagittal and axial planes, respectively.}
    \label{fig7}
\end{figure}

From Table \ref{tab2}, it can be seen that the model with the best overall quantitative results is the $\alpha$-WGANSigmaRat1 model, but after a quick analysis of several scans by some specialists and the authors, the $\alpha$-WGANSigmaRat2 model seems to produce more realistic scans with less blur and fewer structural anomalies, as shown in Figure \ref{fig7}. Therefore, these two models were used for the Turing test. The $\alpha$-WGAN\_ADNI model was discarded for the Turing test because it was too easy to distinguish between real and generated scans.

\begin{table}[h]
\centering
\small
\caption{Quantitative results. For each cell, the first value is the mean and the second is the standard deviation. In bold are the best results for each metric. An up arrow ↑ next to the metric name means that larger values are better and a down arrow ↓ means the opposite}
\begin{tabular}{|l|l|l|l|l|}
\hline
Results         & MS-SSIM                                                   & NCC ↑                                                     & MAE ↓                                                     & MMD ↓                                                        \\ \hline
$\alpha$-WGAN\_ADNI    & \begin{tabular}[c]{@{}l@{}}0.6860 \\ ±0.0066\end{tabular} & \begin{tabular}[c]{@{}l@{}}0.7241\\ ±0.0071\end{tabular}  & \begin{tabular}[c]{@{}l@{}}0.0316 \\ ±0.0004\end{tabular} & \begin{tabular}[c]{@{}l@{}}779.4653 \\ ±27.2016\end{tabular} \\ \hline
$\alpha$-WGANSigmaRat1 & \begin{tabular}[c]{@{}l@{}}\textbf{0.8118} \\ \textbf{±0.0051}\end{tabular} & \begin{tabular}[c]{@{}l@{}}\textbf{0.7887} \\ \textbf{±0.0041}\end{tabular} & \begin{tabular}[c]{@{}l@{}}\textbf{0.0305} \\ \textbf{±0.0004}\end{tabular} & \begin{tabular}[c]{@{}l@{}}\textbf{753.1584} \\ \textbf{±24.8816}\end{tabular} \\ \hline
$\alpha$-WGANSigmaRat2 & \begin{tabular}[c]{@{}l@{}}0.8236 \\ ±0.0056\end{tabular} & \begin{tabular}[c]{@{}l@{}}0.7527 \\ ±0.0037\end{tabular} & \begin{tabular}[c]{@{}l@{}}0.0325 \\ ±0.0003\end{tabular} & \begin{tabular}[c]{@{}l@{}}819.3409 \\ ±20.4437\end{tabular} \\ \hline
\end{tabular}
\label{tab2}
\end{table}

The Turing test was performed by four MRI experts on 50 scans, of which 20 were real and 15+15 were generated by each model, $\alpha$-WGANSigmaRat1 and $\alpha$-WGANSigmaRat2. Only 50 scans were selected because each evaluation is very time-consuming. Therefore, it was agreed with the evaluators that 50 scans were a reasonable number. Also, the experts had to follow some rules, such as: do not open more than one scan at a time; do not change any answer; consider only the original slices (axial plane); do not look at the dataset D\textsubscript{r}\textsuperscript{210} while performing the test. Table \ref{tab3} presents the results of the Turing test and the respective expertise of each expert. For each rater, the first three values are the number of failed answers, e.g. $\alpha$-WGANSigmaRat1=13 means that 13 scans generated by the $\alpha$-WGANSigmaRat1 model were misclassified, and the last value is the number of correct answers. 

\begin{table}[h]
\centering
\small
\caption{Turing test results. The first two values presented in the results are the misclassification of the scans generated by the $\alpha$-WGANSigmaRat1 and $\alpha$-WGANSigmaRat2 models, respectively, the third value is the misclassification of the real scans and the last value is the number of right answers}
\begin{tabular}{|l|llll|}
\hline
Raters        & 1    & 2      & 3   & 4      \\ \hline
Expertise     & High & Medium & Low & Medium \\
WGANSigmaRat1 & 1    & 13     & 2   & 5      \\
WGANSigmaRat2 & 0    & 8      & 4   & 8      \\
Real          & 1    & 16     & 3   & 8      \\
Right Answers & 48   & 13     & 41  & 29     \\ \hline
\end{tabular}

\label{tab3}
\end{table}

The generated scans were also evaluated in a segmentation task to determine if the synthetic data could improve the results and if it performed better than traditional data augmentation. Using a DL model developed by Rodrigues (2018) for Grey Matter (GM), White Matter (WM) and CerebroSpinal Fluid (CSF) segmentation\cite{Rodrigues2018} an experiment was conducted using the following datasets:
D\textsubscript{r}\textsuperscript{174} =  {(x\textsubscript{i}, y\textsubscript{i})}\textsubscript{i=1}\textsuperscript{174};
D\textsubscript{r}\textsuperscript{87} =  {(x\textsubscript{i}, y\textsubscript{i})}\textsubscript{i=1}\textsuperscript{87};
D\textsubscript{s}\textsuperscript{174} =  {(x\textsubscript{i}, y\textsubscript{i})}\textsubscript{i=1}\textsuperscript{174};
D\textsubscript{s}\textsuperscript{87} =  {(x\textsubscript{i}, y\textsubscript{i})}\textsubscript{i=1}\textsuperscript{87};
D\textsubscript{s}\textsuperscript{261} =  {(x\textsubscript{i}, y\textsubscript{i})}\textsubscript{i=1}\textsuperscript{261};
D\textsubscript{s}\textsuperscript{348} =  {(x\textsubscript{i}, y\textsubscript{i})}\textsubscript{i=1}\textsuperscript{348}. 
The number of scans in each dataset is a multiple of 174 since the maximum of labelled real scans i.e. with existing segmentation masks of GM, WM and CSF is the dataset D\textsubscript{r}\textsuperscript{174}. Only the $\alpha$-WGANSigmaRat2 model was used to generate synthetic scans because the raters involved in the Turing test, especially the first one, indicated that the scans generated with this model were more realistic, had less blur and more detail, and could perform better on the segmentation task than the $\alpha$-WGANSigmaRat1 model. The Statistical Parametric Mapping software (SPM12, Welcome Centre for Human Neuroimaging \cite{penny2011statistical}) was used to generate the GM, WM and CSF labels for the generated scans. SPM12 was unable to synthesise the labels of some synthetic scans, so these were discarded and new scans were generated, i.e. the SPM12 software was used in a final step to assess the quality of the generated scans before segmentation. Table \ref{tab4} presents the results of this segmentation task performed on different combinations of the previously mentioned datasets. The Dice result of the first test, i.e. segmentation using only real scans, is the same as in Rodrigues (2018) \cite{Rodrigues2018}. The use of the conventional data augmentation with the dataset  D\textsubscript{r}\textsuperscript{174} =  {(x\textsubscript{i}, y\textsubscript{i})}\textsubscript{i=1}\textsuperscript{174} was also tested (Test 10). 

\begin{table}[h]
\centering
\tiny
\caption{Segmentation of grey matter, white matter, and cerebrospinal fluid dice score, where r refers to real and s refers to synthetic scans. If two datasets are included in the same test, this means that the test was performed with the union of both datasets. The WM, GM and CSF Dice are the dice score of the segmentation of white matter, grey matter, and cerebrospinal fluid, respectively. The best results are in bold}
\begin{tabular}{|l|l|l|l|l|l|l|l|l|l|l|}
\hline
Tests                                                    & Test1 & Test2                                             & Test3                                                & Test4                                              & Test5                                              & Test6 & Test7 & Test8                                             & Test9                                             & Test10 \\ \hline
\begin{tabular}[c]{@{}  l@{}}Data \\ sets\end{tabular}     & D\textsubscript{r}\textsuperscript{174}    & \begin{tabular}[c]{@{}l@{}}D\textsubscript{r}\textsuperscript{174}\\ D\textsubscript{s}\textsuperscript{87}\end{tabular} & \begin{tabular}[c]{@{}l@{}}D\textsubscript{r}\textsuperscript{174}  \\ D\textsubscript{s}\textsuperscript{174}\end{tabular} & \begin{tabular}[c]{@{}l@{}}D\textsubscript{r}\textsuperscript{174}\\ D\textsubscript{s}\textsuperscript{261}\end{tabular} & \begin{tabular}[c]{@{}l@{}}D\textsubscript{r}\textsuperscript{174}\\ D\textsubscript{s}\textsuperscript{348}\end{tabular} & D\textsubscript{s}\textsuperscript{174}    & D\textsubscript{s}\textsuperscript{348}    & \begin{tabular}[c]{@{}l@{}}D\textsubscript{r}\textsuperscript{87}\\ D\textsubscript{s}\textsuperscript{174}\end{tabular} & \begin{tabular}[c]{@{}l@{}}D\textsubscript{r}\textsuperscript{87}\\ D\textsubscript{s}\textsuperscript{348}\end{tabular} & D\textsubscript{r}\textsuperscript{174}     \\ \hline
\begin{tabular}[c]{@{}l@{}}Global   \\ Dice\end{tabular} & 0.8969   & 0.9138                                               & 0.9083                                                  & 0.9078                                                & \textbf{0.9141}                                                & 0.8238   & 0.7646   & 0.8979                                               & 0.8259                                               & 0.8183    \\ \hline
\begin{tabular}[c]{@{}l@{}}GM \\ Dice\end{tabular}       & 0.9381   & \textbf{0.9419}                                               & 0.9384                                                  & 0.9376                                                & 0.9412                                                & 0.8863   & 0.8586   & 0.9316                                               & 0.8863                                               & 0.8856    \\ \hline
\begin{tabular}[c]{@{}l@{}}WM\\ Dice\end{tabular}        & 0.8969   & 0.9077                                               & 0.9037                                                  & 0.9014                                                & \textbf{0.9098}                                                & 0.8202   & 0.7262   & 0.8897                                               & 0.8301                                               & 0.7824    \\ \hline
\begin{tabular}[c]{@{}l@{}}CSF \\ Dice\end{tabular}      & 0.7468   & \textbf{0.8232}                                              & 0.8098                                                  & 0.8170                                                & 0.8180                                                & 0.6095   & 0.4418   & 0.7442                                               & 0.6273                                               & 0.6042    \\ \hline
\end{tabular}
\label{tab4}
\end{table}

\section{Discussion}

The initial evaluation of the models was based on the quantitative results of Table \ref{tab2} to discard the worse models more quickly, e.g. the MS-SSIM value should not be very close to 1, as it is important that the model can generate a large variety of scans. Table \ref{tab2} shows that the $\alpha$-WGANSigmaRat1 model has the second highest MS-SSIM value and the value closest to the real distribution that is higher than the real distribution, which means that the generated scans are more repetitive than the real ones. This is not necessarily a problem, as the value is far from 1.0, but it does not correspond to the real distribution of the sigma dataset of rat brains. This value is even higher for the $\alpha$-WGANSigmaRat2 model, but the generated scans seem to be more realistic after a qualitative visual inspection by the experts. The remaining metrics were too close to each other, so it was necessary to use other tools to further compare these two models.

\begin{figure}[h]
    \centering
    \includegraphics[width=1\textwidth]{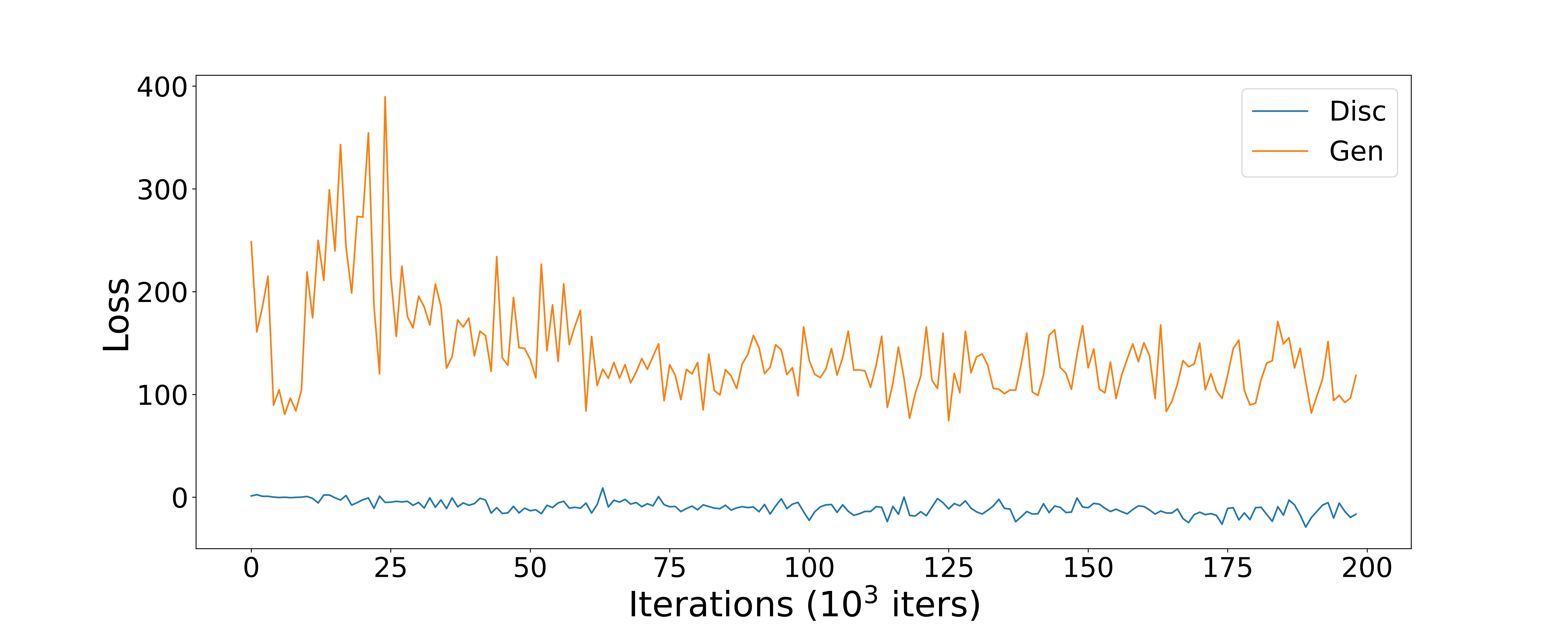}
    \caption{Discriminator (blue) and Generator (orange) loss function plots of the $\alpha$-WGANSigmaRat2 model training process.}
    \label{fig8}
\end{figure}

In the loss function plots of the $\alpha$-WGANSigmaRat2 model (Figure \ref{fig8}), it can be seen that the training of the discriminator was very stable and after 60000 iterations, the training of the generator also stabilised. 

\begin{figure}[h]
    \centering
    \includegraphics[width=1\textwidth]{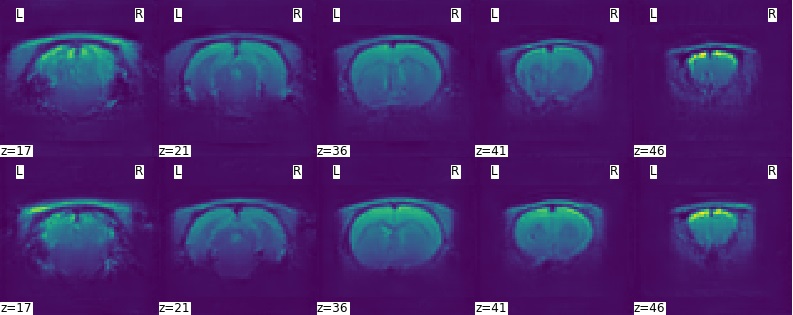}
    \caption{Axial slices of a generated scan with the $\alpha$-WGANSigmaRat2 model after 100000 iterations (first row) and after 200000 iterations (second row) of training.}
    \label{fig9}
\end{figure}

It is not good practice to compare different generative models using loss function diagrams, as they do not correspond to human perception. These diagrams are only a good tool to check for the presence of mode collapse, which is characterised by the divergence between the discriminator and the generator loss function diagrams, one tending to -$\infty$ and the other to +$\infty$. Normally the training runs until this divergence occurs, but since an $\alpha$-GAN architecture with VAE was used, this would never happen or it would require many more iterations, so it was decided to run 200000 iterations. Comparison of the scans generated after 100000 and 200000 training iterations confirmed that the learning process is not directly related to the representation of the loss functions. This can be seen in Figure \ref{fig9} by the lack of detail in the generated scans after 100000 training iterations (first row) and a significant improvement after 200000 training iterations (second row).

The results presented in Table \ref{tab3} show that some experts had several difficulties in distinguishing between real and fake scans, although some of them had no problems due to their experience with MRI acquisition and visualisation of rat brains. Rater 1, who was very familiar with MRIs of rat brains, said that the generated scans looked very grainy, without sufficient anatomical detail, and that the signal distribution outside the skull did not look right. He was the expert best able to distinguish between synthetic and real scans, so his input and comments were very important. Raters 3 and 4 reported some distortions, some strange blurs and low signal homogeneity. Rater 2 is used to analyse MRI scans from humans but not from rats. This could be the reason why he rejected many responses because they were too homogeneous and symmetrical.

\begin{figure}[h]
    \centering
    \includegraphics[width=1\textwidth]{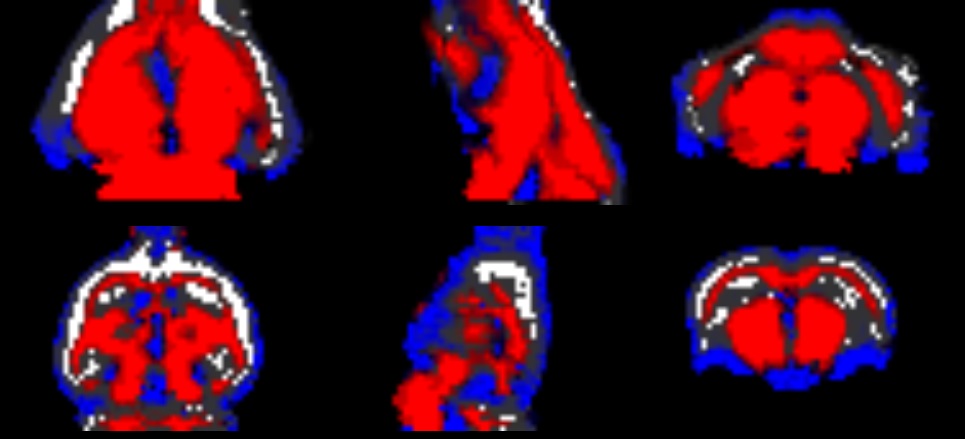}
    \caption{Semantic masks built with the SPM12 software. The semantic mask in the first row was built using a scan generated with the $\alpha$-WGAN\_ADNI model, and the semantic mask in the second row was built using a scan generated with the $\alpha$-WGANSigmaRat2 model. The blue colour is the CSF, the red colour is the WM and the grey scale is the grey matter.}
    \label{fig10}
\end{figure}

After generating the scans for the final test, a selection was made to ensure that all scans had correct anatomy, i.e. whether the SPM12 software could (or could not) generate the GM, WM and CSF labels. About half of the randomly generated scans had structural defects and the SPM software was not able to create the GM, WM and CSF labels correctly. It was also tested whether SPM could create the labels of the synthetic data generated by the remaining models. It was found that almost all the scans generated by $\alpha$-WGAN\_ADNI had problems in creating the labels, so the proposed methods, i.e. the new loss functions and the normalisation layer were able to improve the quality of the generated scans, as can be seen in  Figure \ref{fig10}. 

The segmentation results are shown in Table \ref{tab4}. The best results for global and white matter segmentation were obtained when a dataset combining D\textsubscript{r}\textsuperscript{174} and D\textsubscript{s}\textsuperscript{348} was used, with an improvement in the Dice score of 0.0172 and 0.0129, respectively. For grey matter and CSF segmentation, the use of the D\textsubscript{r}\textsuperscript{174} and the D\textsubscript{s}\textsuperscript{87} resulted in an improvement of 0.0038 and 0.0764, respectively. The introduction of only 87 synthetic scans (D\textsubscript{s}\textsuperscript{87}) into the D\textsubscript{r}\textsuperscript{174} dataset had a big impact, particularly on CSF segmentation, evidencing that the use of GANs can significantly improve results without the need for more scanning sessions. It was also demonstrated that the use of conventional data augmentation (Test 10) was detrimental to the training process, so data augmentation with synthetic data was superior in this case. However, it should be noted that using only synthetic data was harmful, so real scans are mandatory.

\section{Conclusion }

In this work, improved alpha-GAN architectures were proposed to generate synthetic MRI scans of the rat brain. The new proposed loss functions used to train the generator and the new proposed normalisation layer applied to the discriminator enabled more realistic results on rat MRI data. The use of traditional data augmentation also helped to generate more diverse and realistic scans. Different techniques were used to evaluate the different architectures and strategies. The $\alpha$-WGANSigmaRat1 model performed better than the others in the quantitative metrics. However, with the help of some MRI experts, it was determined that the best qualitative model was $\alpha$-WGANSigmaRat2. The Turing test proved that it is possible to trick experts with moderate or low expertise, but it is very difficult to trick experts who are familiar with these datasets. Unfortunately, it was not possible to use this test to check which model generated the best scans. Brain structure segmentation also improved when synthetic scans were added to the original dataset, with the global Dice score improving from 0.8969 to 0.9141. The improvements in CSF segmentation were even more significant, from 0.7468 to 0.8232. It was also found that using synthetic data generated by the new data augmentation model produced better than traditional data augmentation. 

This work has shown that it is possible to adapt tools developed mainly for processing human MR images to preclinical research. From a methodological point of view, it was possible to extend methods that work with 2D images to 3D images, and it was shown that it is possible to obtain improved segmentation results by adding synthetic scans to the original dataset. The use of data augmentation techniques in the context of preclinical research is interesting, particularly the ability to create larger datasets without scanning more animals, which contributes to the ethical 3R rule, i.e. the reducing part in particular. It is important to note that using GANs for data augmentation is not a substitute for traditional data augmentation but a complement. Both together contribute to increasing the amount of data.

\section{Acknowledgements}
The MRI dataset of rat brains was acquired in the context of the Sigma project with the reference FCT-ANR/NEU-OSD/0258/2012. This project was co-financed by the French public funding agency ANR (Agence Nationale pour la Recherche, APP Blanc International II 2012), the Portuguese FCT (Fundação para a Ciência e Tecnologia) and the Portuguese North Regional Operational Program (ON.2-O Novo Norte) under the National Strategic Reference Framework (QREN), through the European Regional Development Fund (FEDER), as well as the Projecto Estratégico cofunded by FCT (PEst-C/SAU/LA0026-/2013) and the European Regional Development Fund COMPETE (FCOMP-01-0124-FEDER-037298). France Life Imaging is acknowledged for its support in funding the NeuroSpin platform of preclinical MRI scanners. This work of André Ferreira and Victor Alves has been supported by FCT- Fundação para a Ciência e a Tecnologia within the R\&D Units Project Scope: UIDB/00319/2020.


\bibliography{elsarticle-template}

\begin{thebibliography}{10}
\expandafter\ifx\csname url\endcsname\relax
  \def\url#1{\texttt{#1}}\fi
\expandafter\ifx\csname urlprefix\endcsname\relax\def\urlprefix{URL }\fi
\expandafter\ifx\csname href\endcsname\relax
  \def\href#1#2{#2} \def\path#1{#1}\fi

\bibitem{denic2011mri}
A.~Denic, S.~I. Macura, P.~Mishra, J.~D. Gamez, M.~Rodriguez, I.~Pirko, {MRI in
  rodent models of brain disorders}, Neurotherapeutics 8~(1) (2011) 3--18.

\bibitem{Brockmann2007}
M.~A. Brockmann, A.~Kemmling, C.~Groden, {Current issues and perspectives in
  small rodent magnetic resonance imaging using clinical MRI scanners} 43
  (2007) 79--87.
\newblock \href {https://doi.org/10.1016/j.ymeth.2007.07.001}
  {\path{doi:10.1016/j.ymeth.2007.07.001}}.

\bibitem{Barriere2019}
D.~A. Barri{\`{e}}re, R.~Magalh{\~{a}}es, A.~Novais, P.~Marques, E.~Selingue,
  F.~Geffroy, F.~Marques, J.~Cerqueira, J.~C. Sousa, F.~Boumezbeur,
  M.~Bottlaender, T.~M. Jay, A.~Cachia, N.~Sousa, S.~M{\'{e}}riaux,
  \href{http://dx.doi.org/10.1038/s41467-019-13575-7}{{The SIGMA rat brain
  templates and atlases for multimodal MRI data analysis and visualization}},
  Nature Communications~(2019) (2019) 1--13.
\newblock \href {https://doi.org/10.1038/s41467-019-13575-7}
  {\path{doi:10.1038/s41467-019-13575-7}}.
\newline\urlprefix\url{http://dx.doi.org/10.1038/s41467-019-13575-7}

\bibitem{Magalhaes2018}
R.~J. d.~S. Magalh{\~{a}}es, {An imaging characterization of the adaptive and
  maladaptive response to chronic stress}, Ph.D. thesis, University of Minho
  (2018).

\bibitem{Magalhaes2018a}
R.~Magalh{\~{a}}es, D.~A. Barri{\`{e}}re, A.~Novais, F.~Marques, P.~Marques,
  J.~Cerqueira, J.~C. Sousa, A.~Cachia, F.~Boumezbeur, M.~Bottlaender, T.~M.
  Jay, S.~M{\'{e}}riaux, N.~Sousa,
  \href{http://dx.doi.org/10.1038/mp.2017.244}{{The dynamics of stress: a
  longitudinal MRI study of rat brain structure and connectome}}, Molecular
  Psychiatry 23~(10) (2018) 1998--2006.
\newblock \href {https://doi.org/10.1038/mp.2017.244}
  {\path{doi:10.1038/mp.2017.244}}.
\newline\urlprefix\url{http://dx.doi.org/10.1038/mp.2017.244}

\bibitem{Magalhaes2019}
R.~Magalh{\~{a}}es, E.~Ganz, M.~Rodrigues, D.~A. Barri{\`{e}}re,
  S.~M{\'{e}}riaux, T.~M. Jay, N.~Sousa, {Biomarkers of resilience and
  susceptibility in rodent models of stress}, Stress Resilience: Molecular and
  Behavioral Aspects~(November) (2019) 311--321.
\newblock \href {https://doi.org/10.1016/B978-0-12-813983-7.00020-3}
  {\path{doi:10.1016/B978-0-12-813983-7.00020-3}}.

\bibitem{Boucher2017}
M.~Boucher, F.~Geffroy, S.~Pr{\'{e}}v{\'{e}}ral, L.~Bellanger, E.~Selingue,
  G.~Adryanczyk-Perrier, M.~P{\'{e}}an, C.~T. Lef{\`{e}}vre, D.~Pignol,
  N.~Ginet, S.~M{\'{e}}riaux, {Genetically tailored magnetosomes used as MRI
  probe for molecular imaging of brain tumor}, Biomaterials 121 (2017)
  167--178.
\newblock \href {https://doi.org/10.1016/j.biomaterials.2016.12.013}
  {\path{doi:10.1016/j.biomaterials.2016.12.013}}.

\bibitem{Vanhoutte2005}
G.~Vanhoutte, I.~Dewachter, P.~Borghgraef, F.~{Van Leuven}, A.~{Van Der
  Linden}, {Noninvasive in vivo MRI detection of neuritic plaques associated
  with iron in APP[V717I] transgenic mice, a model for Alzheimer's disease},
  Magnetic Resonance in Medicine 53~(3) (2005) 607--613.
\newblock \href {https://doi.org/10.1002/mrm.20385}
  {\path{doi:10.1002/mrm.20385}}.

\bibitem{Foroozandeh2020}
M.~Foroozandeh, A.~Eklund, {Synthesizing brain tumor images and annotations by
  combining progressive growing GAN and SPADE}, arXiv preprint arXiv:2009.05946
  (2020).
\newblock \href {http://arxiv.org/abs/arXiv:2009.05946v1}
  {\path{arXiv:arXiv:2009.05946v1}}.

\bibitem{russell1959principles}
W.~M.~S. Russell, R.~L. Burch, {The principles of humane experimental
  technique}, Methuen, 1959.

\bibitem{Nalepa2019}
J.~Nalepa, M.~Marcinkiewicz, M.~Kawulok, {Data Augmentation for Brain-Tumor
  Segmentation: A Review}, Frontiers in Computational Neuroscience
  13~(December) (2019) 1--18.
\newblock \href {https://doi.org/10.3389/fncom.2019.00083}
  {\path{doi:10.3389/fncom.2019.00083}}.

\bibitem{Sandfort2019}
V.~Sandfort, K.~Yan, P.~J. Pickhardt, R.~M. Summers,
  \href{http://dx.doi.org/10.1038/s41598-019-52737-x}{{Data augmentation using
  generative adversarial networks (CycleGAN) to improve generalizability in CT
  segmentation tasks}}, Scientific Reports 9~(1) (2019) 1--9.
\newblock \href {https://doi.org/10.1038/s41598-019-52737-x}
  {\path{doi:10.1038/s41598-019-52737-x}}.
\newline\urlprefix\url{http://dx.doi.org/10.1038/s41598-019-52737-x}

\bibitem{Motamed2020}
S.~Motamed, P.~Rogalla, F.~Khalvati,
  \href{http://arxiv.org/abs/2006.03622}{{Data Augmentation using Generative
  Adversarial Networks (GANs) for GAN-based Detection of Pneumonia and COVID-19
  in Chest X-ray Images}} (2020) 1--12\href {http://arxiv.org/abs/2006.03622}
  {\path{arXiv:2006.03622}}, \href {https://doi.org/10.21203/rs.3.rs-146161/v1}
  {\path{doi:10.21203/rs.3.rs-146161/v1}}.
\newline\urlprefix\url{http://arxiv.org/abs/2006.03622}

\bibitem{el2019deep}
M.~El-Kaddoury, A.~Mahmoudi, M.~M. Himmi, {Deep generative models for image
  generation: A practical comparison between variational autoencoders and
  generative adversarial networks}, in: International Conference on Mobile,
  Secure, and Programmable Networking, Springer, 2019, pp. 1--8.

\bibitem{Goodfellow2014}
I.~J. Goodfellow, J.~Pouget-Abadie, M.~Mirza, B.~Xu, D.~Warde-Farley, S.~Ozair,
  A.~Courville, Y.~Bengio, {Generative adversarial nets}, Advances in Neural
  Information Processing Systems 3~(January) (2014) 2672--2680.
\newblock \href {http://arxiv.org/abs/arXiv:1406.2661v1}
  {\path{arXiv:arXiv:1406.2661v1}}.

\bibitem{Alqahtani2019}
H.~Alqahtani, M.~Kavakli-Thorne, G.~Kumar,
  \href{https://doi.org/10.1007/s11831-019-09388-y}{{Applications of Generative
  Adversarial Networks (GANs): An Updated Review}}, Archives of Computational
  Methods in Engineering~(0123456789) (2019).
\newblock \href {https://doi.org/10.1007/s11831-019-09388-y}
  {\path{doi:10.1007/s11831-019-09388-y}}.
\newline\urlprefix\url{https://doi.org/10.1007/s11831-019-09388-y}

\bibitem{Gui2020}
J.~Gui, Z.~Sun, Y.~Wen, D.~Tao, J.~Ye,
  \href{http://arxiv.org/abs/2001.06937}{{A Review on Generative Adversarial
  Networks: Algorithms, Theory, and Applications}} 14~(8) (2020) 1--28.
\newblock \href {http://arxiv.org/abs/2001.06937} {\path{arXiv:2001.06937}}.
\newline\urlprefix\url{http://arxiv.org/abs/2001.06937}

\bibitem{Brock2019}
A.~Brock, J.~Donahue, K.~Simonyan, {Large scale GaN training for high fidelity
  natural image synthesis}, 7th International Conference on Learning
  Representations, ICLR 2019 (2019) 1--35\href
  {http://arxiv.org/abs/arXiv:1809.11096v2} {\path{arXiv:arXiv:1809.11096v2}}.

\bibitem{Karras2020}
T.~Karras, S.~Laine, M.~Aittala, J.~Hellsten, J.~Lehtinen, T.~Aila, {Analyzing
  and improving the image quality of stylegan}, Proceedings of the IEEE
  Computer Society Conference on Computer Vision and Pattern Recognition (2020)
  8107--8116\href {http://arxiv.org/abs/1912.04958} {\path{arXiv:1912.04958}},
  \href {https://doi.org/10.1109/CVPR42600.2020.00813}
  {\path{doi:10.1109/CVPR42600.2020.00813}}.

\bibitem{shin2018medical}
H.-C. Shin, N.~A. Tenenholtz, J.~K. Rogers, C.~G. Schwarz, M.~L. Senjem, J.~L.
  Gunter, K.~P. Andriole, M.~Michalski, {Medical image synthesis for data
  augmentation and anonymization using generative adversarial networks}, in:
  International workshop on simulation and synthesis in medical imaging,
  Springer, 2018, pp. 1--11.

\bibitem{NEURIPS2019_9015}
A.~Paszke, S.~Gross, F.~Massa, A.~Lerer, J.~Bradbury, G.~Chanan, T.~Killeen,
  Z.~Lin, N.~Gimelshein, L.~Antiga, A.~Desmaison, A.~Kopf, E.~Yang, Z.~DeVito,
  M.~Raison, A.~Tejani, S.~Chilamkurthy, B.~Steiner, L.~Fang, J.~Bai,
  S.~Chintala,
  \href{http://papers.neurips.cc/paper/9015-pytorch-an-imperative-style-high-performance-deep-learning-library.pdf}{{PyTorch:
  An Imperative Style, High-Performance Deep Learning Library}}, in: Advances
  in Neural Information Processing Systems 32, Curran Associates, Inc., 2019,
  pp. 8024--8035.
\newline\urlprefix\url{http://papers.neurips.cc/paper/9015-pytorch-an-imperative-style-high-performance-deep-learning-library.pdf}

\bibitem{Monai}
T.~M. Consortium, {Project MONAI.} (2020).
\newblock \href {https://doi.org/10.5281/zenodo.4323059}
  {\path{doi:10.5281/zenodo.4323059}}.

\bibitem{Kwon2019}
G.~Kwon, C.~Han, D.~shik Kim, {Generation of 3D Brain MRI Using Auto-Encoding
  Generative Adversarial Networks}, Lecture Notes in Computer Science
  (including subseries Lecture Notes in Artificial Intelligence and Lecture
  Notes in Bioinformatics) 11766 LNCS (2019) 118--126.
\newblock \href {http://arxiv.org/abs/1908.02498} {\path{arXiv:1908.02498}},
  \href {https://doi.org/10.1007/978-3-030-32248-9_14}
  {\path{doi:10.1007/978-3-030-32248-9_14}}.

\bibitem{Kingma2015}
D.~P. Kingma, J.~L. Ba, {Adam: A method for stochastic optimization}, 3rd
  International Conference on Learning Representations, ICLR 2015 - Conference
  Track Proceedings (2015) 1--15\href {http://arxiv.org/abs/1412.6980}
  {\path{arXiv:1412.6980}}.

\bibitem{Sun2020}
Y.~Sun, P.~Yuan, Y.~Sun, {MM-GAN: 3D MRI data augmentation for medical image
  segmentation via generative adversarial networks}, Proceedings - 11th IEEE
  International Conference on Knowledge Graph, ICKG 2020 (2020) 227--234\href
  {https://doi.org/10.1109/ICBK50248.2020.00041}
  {\path{doi:10.1109/ICBK50248.2020.00041}}.

\bibitem{Miyato2018}
T.~Miyato, T.~Kataoka, M.~Koyama, Y.~Yoshida, {Spectral normalization for
  generative adversarial networks}, 6th International Conference on Learning
  Representations, ICLR 2018 - Conference Track Proceedings (2018).
\newblock \href {http://arxiv.org/abs/1802.05957} {\path{arXiv:1802.05957}}.

\bibitem{Agrawal2019}
N.~Agrawal, R.~Katna,
  \href{http://link.springer.com/10.1007/978-981-13-6772-4}{{Applications of
  Computing, Automation and Wireless Systems in Electrical Engineering}}, Vol.
  553, Springer Singapore, 2019.
\newblock \href {https://doi.org/10.1007/978-981-13-6772-4}
  {\path{doi:10.1007/978-981-13-6772-4}}.
\newline\urlprefix\url{http://link.springer.com/10.1007/978-981-13-6772-4}

\bibitem{Mathieu2016}
M.~Mathieu, C.~Couprie, Y.~LeCun, {Deep multi-scale video prediction beyond
  mean square error}, 4th International Conference on Learning Representations,
  ICLR 2016 - Conference Track Proceedings~(2015) (2016) 1--14.
\newblock \href {http://arxiv.org/abs/1511.05440} {\path{arXiv:1511.05440}}.

\bibitem{Chen2018}
Y.~Chen, F.~Shi, A.~G. Christodoulou, Y.~Xie, Z.~Zhou, D.~Li, {Efficient and
  accurate MRI super-resolution using a generative adversarial network and 3D
  multi-level densely connected network}, Lecture Notes in Computer Science
  (including subseries Lecture Notes in Artificial Intelligence and Lecture
  Notes in Bioinformatics) 11070 LNCS (2018) 91--99.
\newblock \href {http://arxiv.org/abs/1803.01417} {\path{arXiv:1803.01417}},
  \href {https://doi.org/10.1007/978-3-030-00928-1_11}
  {\path{doi:10.1007/978-3-030-00928-1_11}}.

\bibitem{Sanchez2018}
I.~S{\'{a}}nchez, V.~Vilaplana, {Brain MRI super-resolution using 3D generative
  adversarial networks}, arXiv~(Midl) (2018) 1--8.
\newblock \href {http://arxiv.org/abs/1812.11440} {\path{arXiv:1812.11440}}.

\bibitem{Loshchilov2019}
I.~Loshchilov, F.~Hutter, {Decoupled weight decay regularization}, 7th
  International Conference on Learning Representations, ICLR 2019 (2019).
\newblock \href {http://arxiv.org/abs/1711.05101} {\path{arXiv:1711.05101}}.

\bibitem{Rodrigues2018}
M.~F. Rodrigues, \href{http://hdl.handle.net/1822/64177}{{Brain Semantic
  Segmentation: A DL approach in Human and Rat MRI studies}}~(November) (2018).
\newline\urlprefix\url{http://hdl.handle.net/1822/64177}

\bibitem{penny2011statistical}
W.~D. Penny, K.~J. Friston, J.~T. Ashburner, S.~J. Kiebel, T.~E. Nichols,
  {Statistical parametric mapping: the analysis of functional brain images},
  Elsevier, 2011.

\end{thebibliography}

\end{document}